# BEAM-BEAM STUDIES FOR THE HIGH-ENERGY LHC

K. Ohmi, KEK, Tsukuba, Japan; O. Dominguez, F. Zimmermann, CERN, Geneva, Switzerland

*Abstract*

LHC upgrades are being considered both towards higher luminosity (HL-LHC) and towards higher energy (HE-LHC). In this paper we report initial studies of the beam-beam effects in the HE-LHC [1]. The HE-LHC aims at beam energies of 16.5 TeV, where the transverse emittance decreases due to synchrotron radiation with a 2-hour damping time. As a result of this emittance, shrinkage the beam-beam parameter increases with time, during a physics store. The beam-beam limit in the HE-LHC is explored using computer simulations.

## INTRODUCTION

At the High-Energy LHC, the proton beam energy is increased from the present LHC design energy of 7 TeV to an upgrade value of 16.5 TeV. The HE-LHC target luminosity is $2 \times 10^{34}$ cm$^{-2}$ s$^{-1}$. The parameters of the nominal and the High-Energy LHC are summarized in Table 1. The radiation damping time, which is 1 and 2 hours for the longitudinal and transverse plane, respectively, will be visible in operation.

The equilibrium horizontal emittance and energy spread that would arise from a balance of radiation damping and random quantum excitation (as in a typical electron storage ring) are very small, $\varepsilon_x \sim 5 \times 10^{-12}$ m and $\sigma_p \sim 1.1 \times 10^{-5}$, respectively. At the quantum equilibrium the diffusion rates per turn of the quantum radiation excitation are $\langle \Delta\varepsilon_x \rangle = 5.8 \times 10^{-20}$ m and $\langle \Delta\varepsilon_z \rangle = 4.2 \times 10^{-15}$ m, respectively, in the absence of any additional blowup. Assuming 20% emittance coupling, the vertical emittance and diffusion rate are $\varepsilon_y = 1 \times 10^{-12}$ m and $\langle \Delta\varepsilon_y \rangle = 1.2 \times 10^{-20}$ m respectively.

Table 1 Parameter list of nominal and high-energy LHC

|  | nominal | HE-LHC |
|---|---|---|
| Beam Energy (TeV) | 7 | 16.5 |
| Bunch population | $1.15 \times 10^{11}$ | $1.29 \times 10^{11}$ |
| Emittance $x/y$ (m) | $5.1 \times 10^{-10}$ | $2.1/1.0 \times 10^{-10}$ |
| Bunch length (m) | 0.0755 | 0.065 |
| Energy spread ($10^{-4}$) | 1.13 | 0.9 |
| $\beta^*$ $x/y$ (m) | 0.55/0.55 | 1/0.43 |
| Damping time x&y/z (h) | 25.8/12.9† | 1.97/0.98† |
| Number of bunches | 2808 | 1404 |
| Luminosity (cm$^{-2}$ s$^{-1}$) | $1.0 \times 10^{34}$ | $2.0 \times 10^{34}$ |

†Here the damping time refers to the emittance decrease, $\varepsilon_i = \varepsilon_{0,i} \exp(-t/\tau_i)$, not to amplitude. The amplitude damping times would be two times longer.

The quantum equilibrium is not reached, however, since intra-beam scattering (IBS) also causes a random excitation of the beam. The IBS diffusion rate depends on the phase space volume of the beam. The diffusion rate due to intra-beam scattering can be estimated using the nominal LHC optics and the MADX IBS module. The emittance growth rates found in this way are 64, 400 and 80 hours for the horizontal, vertical and longitudinal plane, respectively, at the initial design emittance. The diffusion rates per turn translate to $\langle \Delta\varepsilon_x \rangle = 6.6 \times 10^{-20}$ m, $\langle \Delta\varepsilon_y \rangle = 6.5 \times 10^{-21}$ m and $\langle \Delta\varepsilon_z \rangle = 2.3 \times 10^{-15}$ m. Therefore, initially the transverse rates are comparable to the radiation excitation. The diffusion rates of the intra-beam scattering strongly increase for smaller beam emittances. The equilibrium emittances reached due to the interplay of intra-beam diffusion and radiation damping are calculated as $8.6 \times 10^{-11}$ m, $1.7 \times 10^{-11}$ m and $1.4 \times 10^{-6}$ m. At this IBS equilibrium, the diffusion rates per turn are $1.1 \times 10^{-18}$ m, $4.2 \times 10^{-20}$ m and $3.5 \times 10^{-15}$ m. The beam-beam parameters for these emittances and with the initial bunch charge are 0.018 ($x$) and 0.025 ($y$). Keeping the longitudinal emittance constant, equal to 2.5 eVs, by means of an external excitation, the transverse equilibrium emittances become $5.1 \times 10^{-11}$ m ($x$) and $1.0 \times 10^{-11}$ m ($y$), resulting in the diffusion rates per turn of $6.2 \times 10^{-19}$ m ($x$) and $2.5 \times 10^{-20}$ m ($y$). Figure 1 shows the evolution of the emittance as a function of time. The beam-beam parameters for the final emittances are 0.026/IP ($x$) and 0.037/IP ($y$). In equilibrium with the radiation damping, the diffusion rate always equals $\langle \Delta\varepsilon_i \rangle = \varepsilon_i T_0 / \tau_I$, or $\langle \Delta x^2 \rangle^{1/2} = (T_0/\tau_x)^{1/2} \sigma_{x,eq} = 1.1 \times 10^{-4} \sigma_{x,eq}$.

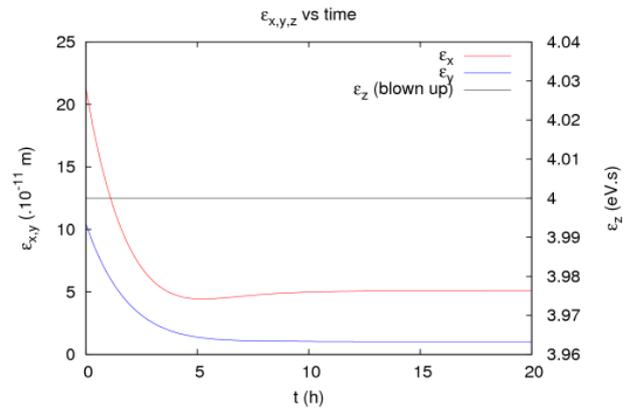

Figure 1: Horizontal and vertical emittances as functions of time. The radiation damping and the diffusion due to intra-beam scattering are taken into account, while proton burn off is not. It is assumed that the longitudinal emittance is continually blown up, so as to acquire a constant value of 4 eVs (=$4\pi\sigma_E\sigma_t$).

In this report, we do not take into account the proton loss due to the collision, partly since the bunch population and the operation scheme are not yet fixed for HE-LHC. Instead the beam-beam limit related to the damping of the emittance is an interesting subject for us. In the beam-beam simulation, diffusion with a constant rate is taken into account in each plane. Our diffusion model does not represent the exact IBS diffusion. This approximation is justified if the emittances and diffusion rates do not vary enormously during the course of the simulation.

We discuss coherent and incoherent effects due to the beam-beam interaction by considering the high geometrical beam-beam parameters 0.026/IP ($x$) and 0.037/IP ($y$), and the aforementioned corresponding values for the IBS diffusion rates.

## COHERENT EFFECT IN HE-LHC

In this section, we discuss coherent beam-beam effect using the strong-strong beam-beam simulation code BBSS. We consider a single interaction point. The collision is simulated by 2D model: i.e., the crossing angle is not taken into account. The actual diffusion rate is very small considering the typical statistics of the simulation using 1 million (0.1%) macro-particles and the resulting numerical noise. The simulations using the real radiation damping and the diffusion rate are hard for the computation time.

We study several model cases with faster damping times (and correspondingly increased diffusion rates) and then try to extrapolate the results to the real case. The damping times are assumed to be either 3.55 or 35.5 s, which is 2000 or 200 times faster than for the HE-LHC. The case with 20 time faster damping time has also been tried, but definite results could not be obtained within an acceptable calculation time.

Figure 2 shows the evolution of the luminosity and the beam size for the damping time of 3.55 s. The difference of the top and bottom two plots is the existence (top) or absence of diffusion (bottom). The diffusion rates of $6 \times 10^{-17}$ ($x$) and $6 \times 10^{-18}$ m ($y$) per turn are taken to be 200 times bigger than the actual IBS diffusion rates (together with the 2000 times faster damping rate leading to a ten times smaller equilibrium emittance), because the natural IBS diffusion rate is smaller than the noise induced by the limited number of macro-particles. The left plots of Fig. 2 display the luminosity per bunch and the beam size. The luminosity increases over the first $15\text{-}17 \times 10^4$ turns, and then drops. At the same time the beam size shrinks up to the same number of turns and then increases. The beam-beam parameters calculated from the beam size, and the dipole amplitudes of the both beam are depicted in the right-hand plots. We can see that the luminosity drop is caused by a coherent dipole-mode beam-beam instability. The beam-beam parameter where the coherent instability arises is quite high, $\xi$=0.15. Already earlier, another weaker coherent instability is seen, after about $8 \times 10^4$ turns, in the top right picture. The beam-beam parameter is 0.03-0.04 at the occurrence of this weak coherent instability. Luminosity degradation is not visible here, and the weak instability disappears after $10^5$ turns. The instability is also seen for the diffusion free case in the right bottom plot, though the amplitude is weaker than on the right top. The diffusion may enhance the coherent motion.

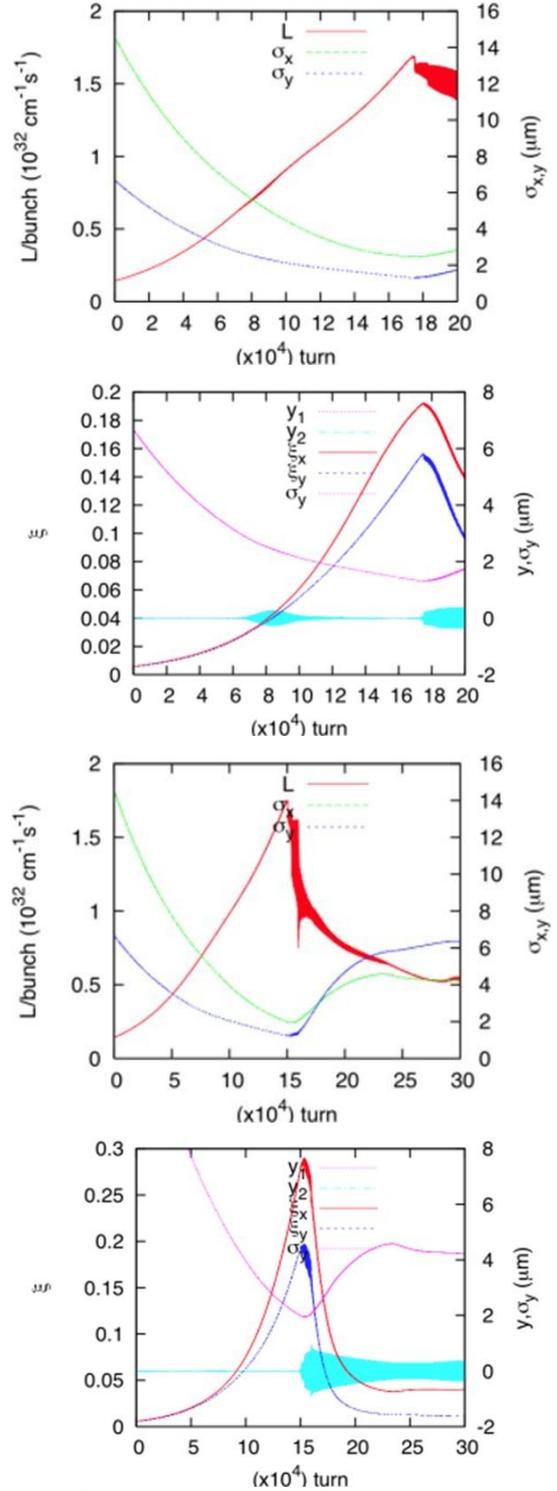

Figure 2: Evolution of the luminosity, beam size, beam-beam parameter and vertical dipole amplitudes. The difference between the top two and bottom two plots is existence (top) or absence of diffusion (bottom). The damping time is assumed to be 3.55 s (4,000 turns).

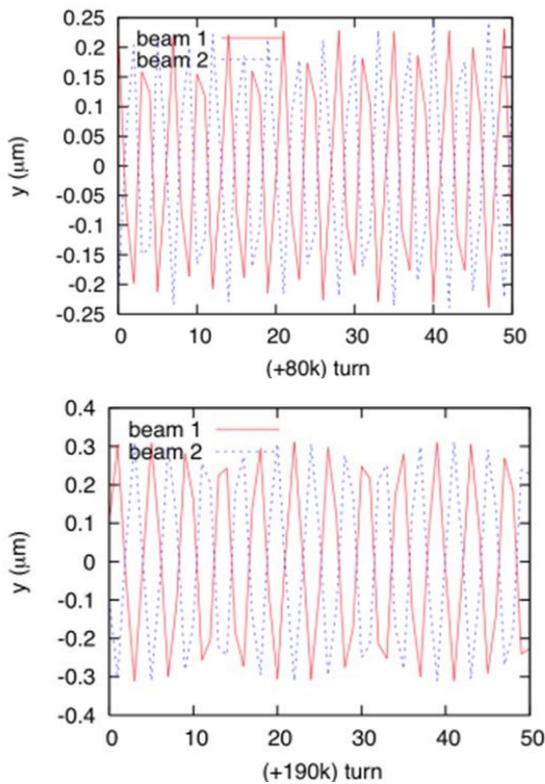
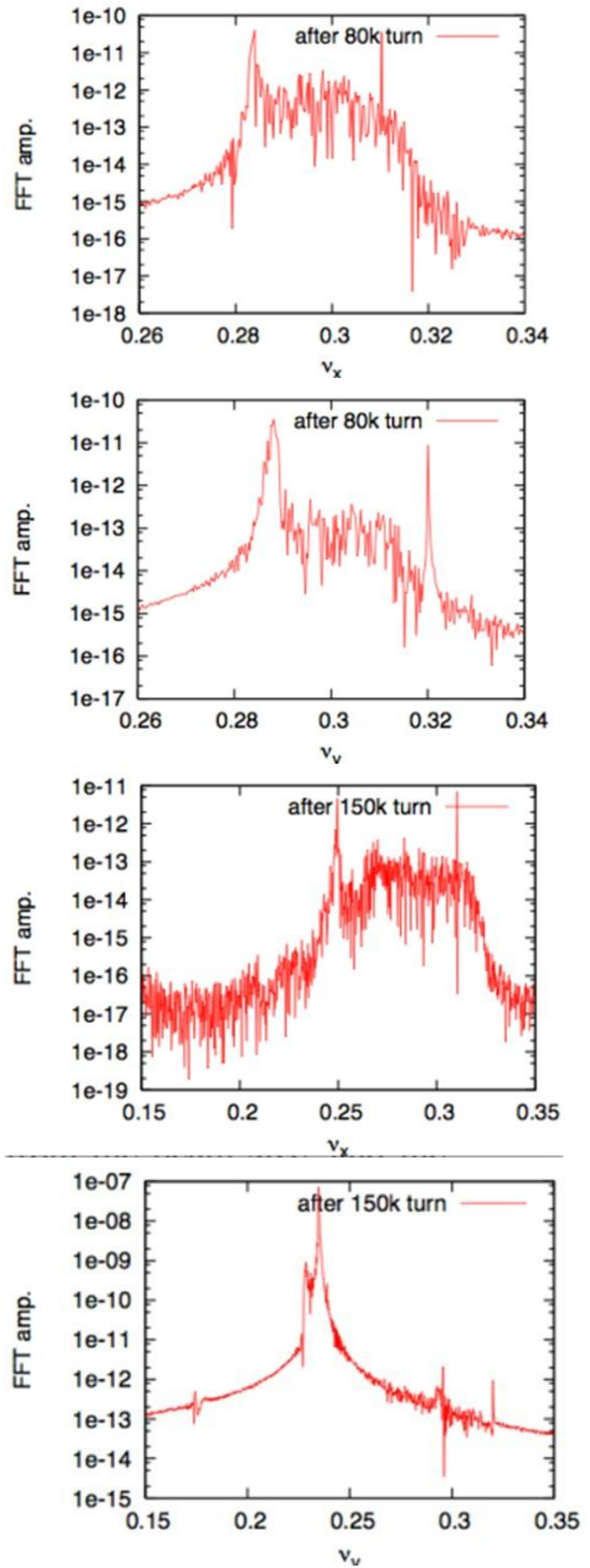

Figure 3: Coherent motion seen in Figure 1. The plots show the dipole amplitudes of the two colliding beams, in the presence of diffusion.

Figures 3 and 4 illustrate the coherent dipole amplitudes of the two colliding beams and Fourier spectrum of the motion for one beam, respectively. We clearly notice a π mode signal in both the beam oscillation and Fourier spectra.

Figure 5 shows the luminosity and beam size for the ten times slower damping time 35.5 s without diffusion. The results roughly scale by a factor ten in time compared with those obtained for the damping time of 3.55 s. The weak instability now occurs around 800,000 turns, though it is hard to see it in the figure. The strong coherent instability appears after about 1,800,000 turns. In view of the good scaling the results may be extrapolated to the case of the real damping time, in a straightforward manner. Actually, an incoherent emittance growth due to the beam-beam interaction dominates for the damping time, as is shown in next section. In addition, the emittance growth from IBS would also limit the beam size. Therefore such high beam-beam parameter is not realized in practice. A simulation with an "IBS" diffusion rate 20 times bigger than the actual one was also attempted, for the 35.5 s damping time. Here the incoherent emittance growth due to beam-beam and IBS dominated, i.e., the emittance did not shrink sufficiently to induce any coherent motion.

Figure 4: Fourier amplitude of the coherent motion for weak and strong instabilities seen after about 80000 and 180000 turns, respectively, in the presence of diffusion. Top and bottom plots display the horizontal and vertical signal, respectively.

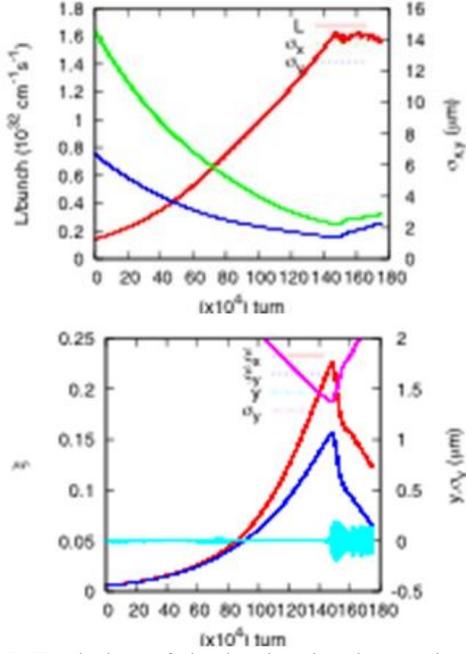

Figure 5: Evolution of the luminosity, beam size, beam-beam parameter and dipole amplitudes for the diffusion-free case. The damping time is assumed to be 35.5 s (40,000 turns).

## INCOHERENT BLOWUP IN HE-LHC

Incoherent emittance growth for the HE-LHC is studied using a weak-strong simulation code (BBWS). Two interaction points are taken into account. A bunch is sliced into 5 pieces along the longitudinal direction. A crossing angle of 170 μrad is taken into account.

A full simulation with the realistic damping time is again too time consuming. Therefore, a weak-strong simulation is performed for various periods of time after starting the collision. We study how a high beam-beam parameter enhances the incoherent emittance growth, in the presence of radiation damping. Table 2 shows the emittance of each stage, in which the simulation is performed. Note that IBS limits the emittance and beam-beam parameters to $(\varepsilon_x, \varepsilon_y)$=(0.051, 0.011) nm and -$(\xi_x, \xi_y)$=(0.026, 0.037), which roughly corresponds the case of $t$=4-5 h. (The equilibrium emittance is realized by overshooting after 10 h in Fig. 1, because the true diffusion rate is a function of emittance.) In the simulation, a smaller emittance (i.e. smaller than the design value of Table 1) is introduced in order to investigate beam-beam emittance growth rate at higher beam-beam parameter.

Figure 6 shows the luminosity evolution at each stage of the beam storage listed in Table 2. The bunch population is kept equal to the initial value. The luminosity for the emittance after $t$=0, 1 and 2 hours (-$\xi$<0.013/IP) does not degrade at all. The luminosity degradation is visible for the emittance after $t \geq 3$ hours (-$\xi$>=0.021/IP). The degradation rate, which is defined as the inverse of luminosity exponential life-time in units of turns, is summarized in Fig. 7. The degradation rates corresponding to luminosity life-times of 1 hour and 1 day, respectively, are depicted in the figure. The beam-beam limit for 1day luminosity life is -$\xi_y$=0.013/IP. Since the damping time is 2 hours for HE-LHC, the limit is -$\xi_y$=0.02/IP.

Table 2: Expected time evolution of emittance and beam-beam parameter for HE-LHC at top energy due to radiation damping, without proton consumption

| t (h) | $\varepsilon_x$ (nm) | $\varepsilon_y$ (nm) | $\xi_x$ (/IP) | $\xi_y$ (/IP) |
|---|---|---|---|---|
| 0 | 0.21 | 0.1 | 0.0051 | 0.0052 |
| 1 | 0.13 | 0.062 | 0.0080 | 0.0084 |
| 2 | 0.076 | 0.037 | 0.012 | 0.013 |
| 3 | 0.046 | 0.022 | 0.017 | 0.021 |
| 4 | 0.027 | 0.014 | 0.023 | 0.031 |
| 5 | 0.016 | 0.0097 | 0.029 | 0.042 |

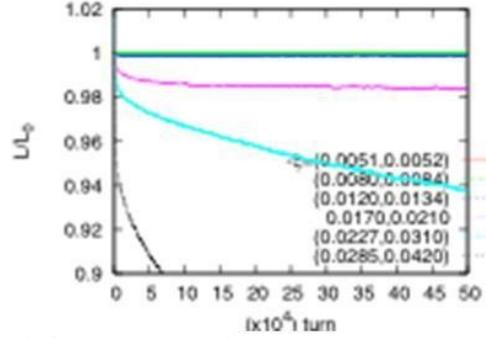

Figure 6: Luminosity evolution assuming the emittance expected after $t$=0-5 hours. The legend is corresponding beam-beam parameters to the emittance.

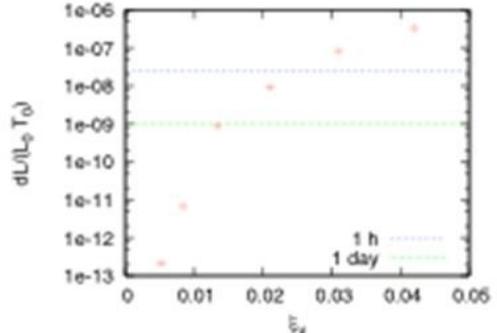

Figure 7: Luminosity degradation rate as function of the vertical beam-beam parameter.

A major source of the luminosity degradation is the crossing angle. Fig. 8 shows the luminosity degradation for collisions without crossing angle (left) and for 285 μrad crossing (right). This simulation was done for the nominal LHC. The luminosity lifetime without crossing angle is 10 times better than the one with the nominal crossing angle. A similar behaviour has also been seen in simulations for KEKB [2].

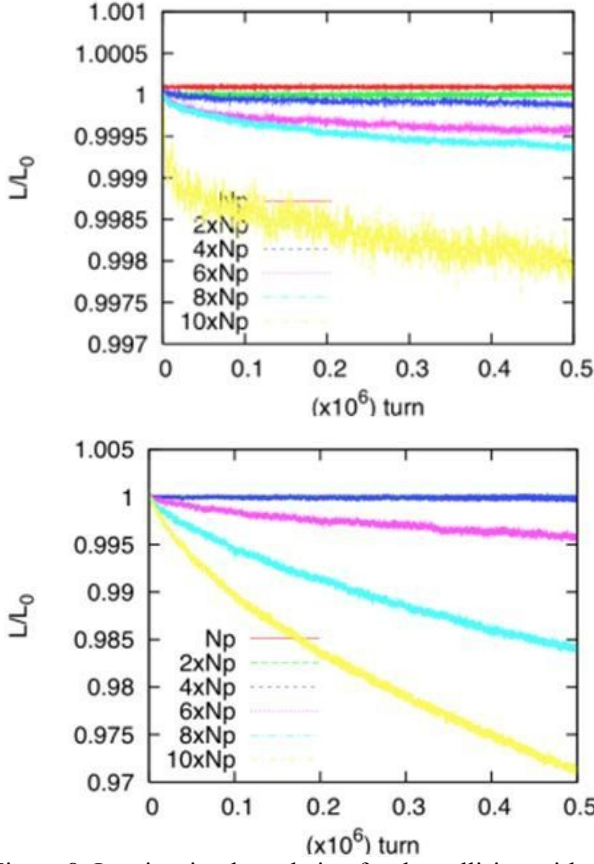

Figure 8: Luminosity degradation for the collision without crossing angle (top) and for 285 μrad crossing (bottom). This simulation was performed for the nominal LHC, with two collision points and alternating crossing.

## EFFECT OF X-Y COUPLING AND DISPERSION FOR THE HE-LHC

In KEKB, the optimization of the linear *x-y* coupling and also of the chromatic coupling at the IP is indispensable to keep a high luminosity during the operation. Tuning of the parameters had continued for 24 hours every day.

The 6x6 revolution matrix, which contains 21 parameters, is parameterized by three sets of Twiss parameters $(\alpha, \beta, \nu)_{xyz}$, four *x-y* coupling parameters, and up to eight dispersion parameters, as follows,

$$M = H^- R^- M_2 R H$$

$$M = \begin{pmatrix} M_x & 0 & 0 \\ 0 & M_z & 0 \\ 0 & 0 & M_z \end{pmatrix}$$

$$M_i = \begin{pmatrix} \cos\mu + \alpha \sin\mu & \beta \sin\mu \\ -\gamma \sin\mu & \cos\mu - \alpha \sin\mu \end{pmatrix}$$

$$R = \begin{pmatrix} r_0 & 0 & r_4 & -r_2 & 0 & 0 \\ 0 & r_0 & -r_3 & r_1 & 0 & 0 \\ -r_1 & -r_2 & r_0 & 0 & 0 & 0 \\ -r_3 & -r_4 & 0 & r_0 & 0 & 0 \\ 0 & 0 & 0 & 0 & 1 & 0 \\ 0 & 0 & 0 & 0 & 0 & 1 \end{pmatrix}$$

$$H = \begin{pmatrix} 1 & 0 & 0 & 0 & 0 & -\eta \\ 0 & 1 & 0 & 0 & 0 & -\eta' \\ 0 & 0 & 1 & 0 & 0 & -\eta_* \\ 0 & 0 & 0 & 1 & 0 & -\eta_*' \\ \eta' & -\eta & \eta_*' & -\eta_* & 1 & 0 \\ 0 & 0 & 0 & 0 & 0 & 1 \end{pmatrix}$$

where four parameters related to $x=\zeta z$ have already been omitted, assuming that there is no transverse kick dependent on z, e.g. no crab cavity, and no cavity placed in a dispersive ($\eta \neq 0$) section. For KEKB the revolution matrix at the collision point determines the collision performance. The tuning performed in KEKB is just a luminosity optimization, performed by scanning r1-r4 and $\eta_y$, $\eta_y$' at the collision point.

We first discuss the effects of the *x-y* coupling on the coherent instability. Figure 9 shows the luminosity (beam-beam parameter) evolution for the case of damping time of 3.5sec. Three lines are given, corresponding to no-coupling, r1=0.01 and 0.05. The threshold for the coherent instability is higher the larger x-y coupling. Perhaps the x-y coupling suppresses the excitation of the coherent mode. However, the suppression is not drastic. The same simulations were done for the other coupling parameters $r_2$-$r_4$. The results were similar: the threshold beam-beam parameter is always higher for larger *x-y* coupling, but the gain is hardly significant.

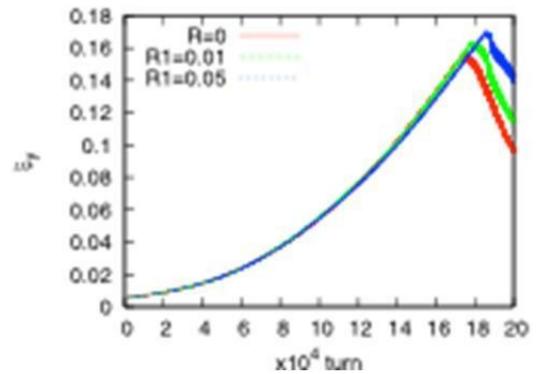

Figure 9: Evolution of beam-beam parameter with or without *x-y* coupling (varying the coupling parameter $r_1$).

Next we discuss the incoherent emittance growth in the presence of *x-y* coupling. Figure 10 shows the luminosity degradation with *x-y* coupling or vertical dispersion. The simulation is performed for a beam-beam parameter of -$\xi_y$=0.02/IP. The 5 plots illustrate the impact of changing $r_1$-$r_4$ and $\eta_y$, respectively. The sensitivity to any of these parameters is quite weak. For KEKB, the tolerances were around $r_1 \sim 0.003$, $r_2 \sim 0.001$, $r_3 \sim r_4 \sim 0.1$. The much reduced sensitivity for the LHC seems to be due to the difference of round (LHC) and flat beams (KEKB).

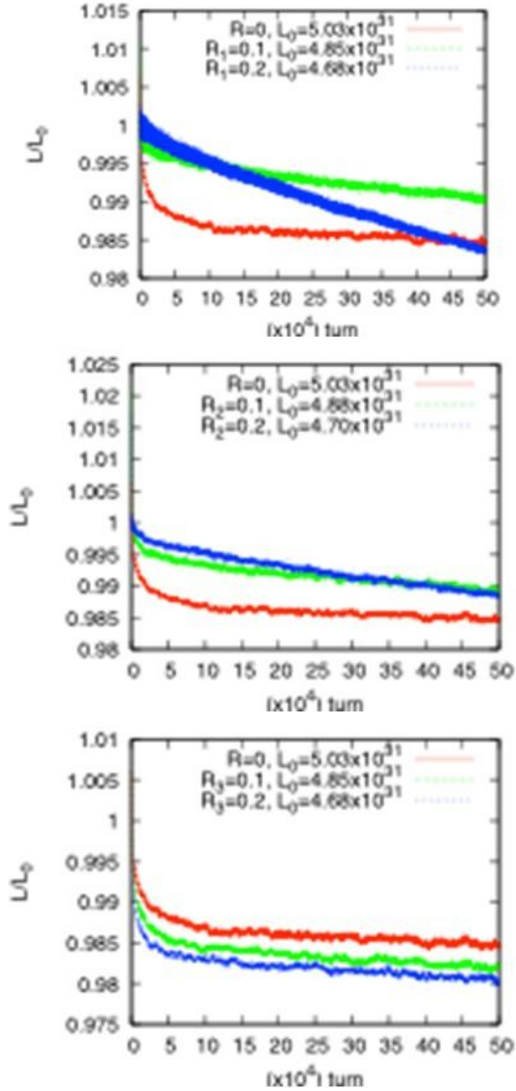

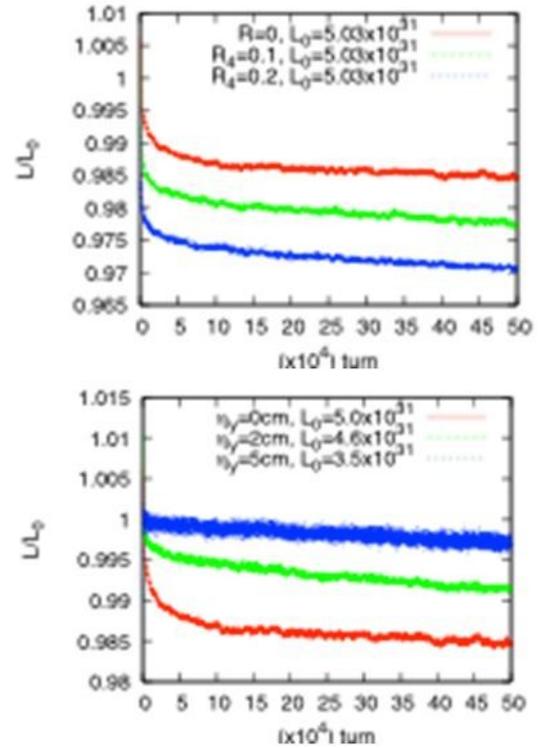

Figure 10: luminosity degradation in the presence of $x$-$y$ coupling or vertical dispersion.

## SUMMARY

Beam-beam effects in the High-Energy LHC have been studied. Both coherent and incoherent phenomena were discussed, using strong-strong and weak-strong simulations, respectively.

A coherent beam-beam instability is induced at high beam-beam parameter $-\xi > 0.15$. The coherent instability is seen in simulations with unrealistically short damping time. For the true damping time, this type of instability is not realized due to the emittance growth caused by the incoherent beam-beam interaction or by IBS.

Incoherent emittance growth was evaluated for several beam-beam parameters. The beam-beam limit is found to be $-\xi = 0.013$/IP without radiation ramping, and $-\xi = 0.02$/IP for a radiation damping time of 2 hours. The incoherent emittance growth is mainly caused by the crossing angle. The emittance growth rate without crossing angle is about 10 times slower.

The sensitivity to $x$-$y$ coupling and spurious vertical dispersion is quite weak compared with the flat-beam collision at the KEK B factory.